\documentclass[prl,aps,twocolumn]{revtex4}
\usepackage{epsfig}
\usepackage{amsmath}
\usepackage{amssymb}
\usepackage{bm}

\newcommand{\beq}{\begin{equation}}
\newcommand{\eeq}{\end{equation}}
\newcommand{\bea}{\begin{eqnarray}}
\newcommand{\eea}{\end{eqnarray}}
\newcommand{\bei}{\begin{itemize}}
\newcommand{\eei}{\end{itemize}}

\newcommand{\br}{{\mathbf{r}}}
\newcommand{\bk}{{\mathbf{k}}}
\def\pf{\mathfrak{Pf}}
\def\Pf#1{\mathfrak{Pf}\left[{#1}\right]}

\def\PLLL{\mathcal{P}_\text{\tiny LLL}}

\def\ket#1{\left|{#1}\right>}

\begin{document}

\title{Paired composite fermion wavefunctions}

\author{G. M\"{o}ller$^1$ and S. H. Simon$^2$}
\affiliation{$^1$ Theory of Condensed Matter Group, Cavendish Laboratory, J.~J.~Thomson Ave., Cambridge CB3~0HE, UK\\
$^2$ Bell Laboratories, Alcatel-Lucent,  Murray Hill, New Jersey 07974}

\begin{abstract}
We construct a family of BCS paired composite fermion
wavefunctions that generalize, but remain in the same topological
phase as, the Moore-Read Pfaffian state for the half-filled Landau
level.
It is shown that for a wide range of experimentally relevant
inter-electron interactions the groundstate can be very accurately
represented in this form.
\end{abstract}
\date{August 20, 2007}
\maketitle

The nature of the fractional quantum Hall effect at
$\nu=\frac{5}{2}$ has been the subject of continued interest since
its discovery roughly two decades ago \cite{Willett87}. The
Moore-Read Pfaffian wavefunction \cite{MooreRead91}, describing
$p$-wave pairing of composite fermions \cite{Greiter92}, is currently
the best candidate wavefunction for this state
\cite{Morf98,HaldaneRezayi00}. Due to the remarkable property of
having quasi-particles with non-abelian statistics, this state has
recently attracted interest in the context of fault-tolerant
topological quantum computation \cite{RevModPhys}. Though the
Moore-Read state is a well established candidate for the groundstate
at $\nu=\frac{5}{2}$, its overlap with the exact groundstate in
simulations of small systems is rather low in comparison to other
known trial states at different filling
factors \cite{Morf98,HaldaneRezayi00}.  This is particularly
discomforting as no explicit construction for perturbations around
the Moore-Read state has been previously available, and the
Moore-Read state has been described as a somewhat unique
choice \cite{Greiter92}. Furthermore several recent studies have
altogether challenged the view of $\nu=\frac{5}{2}$ as being the
Moore-Read state \cite{JainNoPfaffian,JainSecondPaper}.

In this Letter, we introduce a general representation of paired
composite fermion (CF) states, merging the concept of BCS Hall
states \cite{ReadGreen} with the explicit construction of CF
wavefunctions \cite{Heinonen,JainKamilla97}. The Moore-Read state
can be cast very accurately in this form, which reveals its
connection to the pairing of composite fermions on top of a
Fermi-sea, and shows how our general paired CF-BCS wavefunctions are
adiabatically connected to Moore-Read.  We also compare our trial
states to the exact groundstates of the Coulomb Hamiltonian
$\mathcal{H}_C$ for electrons in the 1$^{st}$ excited Landau level
(1LL), plus an arbitrary additional pseudopotential $\delta V_1$
interaction. For a very broad range of $\delta V_1$ we find very
high overlap of our trial wavefunctions with the exact groundstate,
thus showing the extent of the Moore-Read phase.

For our description of the physics at $\nu=\frac{5}{2}$, we shall
assume the lowest Landau level (LLL) to be entirely filled and
inert, such that the relevant degrees of freedom correspond to a
half filled 1LL. We assume the spin degree of freedom of these
electrons to be frozen (although the experimental situation is less
clear \cite{Willett02}). The 1LL is represented by wavefunctions in
the LLL using appropriately modified
pseudopotential coefficients \cite{Haldane83}.

The aim of our construction is to ``composite fermionize'' a simple
BCS state. In second quantized notation, the general form of the BCS
groundstate is \cite{BCS} $\label{eq:BCS} \ket{\Psi_\text{BCS}} =
\prod_\bk \left( 1 + g_\bk \,e^{i\varphi} \, c^\dagger_{\bk}
  c^\dagger_{-\bk}\right)\ket{0},
$ written in an unnormalized manner here. This wavefunction can be
projected to a fixed number of particles by integration over $\int
\mathrm{d}\varphi\,\exp(-iN\varphi)$ such that we retain exactly $N$
pair creation operators. The (inverse) Fourier transform into real
space then yields \cite{BCS} \beq \label{eq:BCSRealSpace}
\Psi_\text{BCS}(\br_1,\ldots,\br_N) = \Pf{g(\br_i-\br_j)}, \eeq
where the Pfaffian $\pf$ is an antisymmetrized sum over all possible
pairings $\pf(g_{ij}) = {\cal A} [g_{12} g_{34} \ldots g_{N-1,N}] =
\pm \sqrt{|\det g_{ij}|}$ with $\cal A$ the antisymmetrization
operator.   In Eq.~\ref{eq:BCSRealSpace}, $g$ is constrained to be
an antisymmetric function, given in terms of its Fourier components
by \beq \label{eq:gFourier} g(\br_i-\br_j) = \sum_\bk g_\bk
e^{i\bk\cdot(\br_i-\br_j)} \equiv \sum_\bk g_\bk \phi_\bk(\br_i)
\phi_{-\bk}(\br_j). \eeq For the last equivalence, we have
identified the exponential factor as the product of two basis
functions $\phi_\bk(\br)=\exp(i\bk \cdot \br)$ of free electrons on
the plane.  This product of free wavefunctions form is naturally
generalized to spherical geometry below.

In order to obtain a LLL wavefunction at filling factor
$\nu=\frac{1}{2}$, we follow Jain's approach \cite{Heinonen} of
multiplying a bare electron wavefunction with Jastrow factors, and
projecting the result to the LLL, yielding \cite{endnote1} \beq
\label{eq:CFPfaffian} \nonumber \Psi^\text{\tiny
CF}_0(z_1,\ldots,z_N) = \PLLL\left\{\Pf{g(\br_i-\br_j)}
\prod_{i<j}(z_i-z_j)^2\right\} \eeq where $\PLLL$ is the LLL
projection operator, and $z_i$ 
is the complex representation of $\br_i$.  The special case  $g =
1/(z_i - z_j)$ reproduces the Moore-Read wavefunction (and the
projection then becomes trivial).

In order to render the projection $\PLLL$ numerically tractable in
general, we bring single particle Jastrow factors $J_i=\prod_{k\neq
i}(z_i-z_k)$ inside the Pfaffian on every line $i$ and every column
$j$ of the matrix $g_{ij}$, and project each of the matrix elements
individually \cite{JainKamilla97}.
As demonstrated in Ref.~\cite{JainKamilla97}, such slight changes in
the projection prescription do not alter the accuracy of the
composite fermionization procedure.   It is thus expected that
projecting matrix elements $[g(\br_i-\br_j) J_i J_j]$ individually
is very similar to a projection of the full wavefunction. For
further simplification, one can decompose $g$ analogous to
Eq.~\ref{eq:gFourier}, and apply the projection separately to each
of the orbitals $\phi_\bk$ as suggested in
Ref.~\cite{JainKamilla97}. Again this slight change in projection
prescription is not expected to damage our wavefunction.   Denoting
$\tilde\phi_\bk(z_i)= J_i^{-1}\PLLL [\phi_\bk(z_i)J_i ]$, Jastrow
factors may be factored again outside the Pfaffian, and we obtain
the final expression for general composite fermionized BCS (CF-BCS)
states
\beq \label{eq:PairedCFState}
\Psi^\text{\tiny CF}=\Pf{
 \sum_\bk g_\bk\, \tilde\phi_\bk(z_i)
\,\tilde\phi_{-\bk}(z_j)}\prod_{i<j}(z_i-z_j)^2. \eeq

In the remainder of this study we will focus on finite size systems
with $N$ electrons on the spherical geometry \cite{Haldane83}.  In
order for Eq.~\ref{eq:PairedCFState} to represent the Moore-Read
phase, we must work at a flux of $N_\phi = 2 N - 3$.  The orbitals
$\phi_\bk$ thus correspond to composite fermions in one quantum of
negative effective flux \cite{Moller05}, ie., the (very small)
effective magnetic field experienced by CF's is directed opposite to
the external magnetic field.  The relevant CF orbitals ($\tilde
\phi_\bk$) are given by the projected monopole harmonics $\tilde
Y^{q=-\frac{1}{2}}_{n,m}$ studied in Ref.~\cite{Moller05}.  To
assure \cite{ReadGreen} that the angular momentum of a pair is $l=-1$
(the negative $p$-wave pairing of the Moore-Read phase) we must
choose $g_\bk \to (-1)^{q+m} g_n$ and we are left with only one
variational parameter $g_n$ per CF shell.  Thus the term in the
brackets of Eq.~\ref{eq:PairedCFState} becomes $\sum_{n,m}
(-1)^{m-1/2} g_n \tilde Y^{q=-\frac{1}{2}}_{n,m}(z_i) \tilde
Y^{q=-\frac{1}{2}}_{n,-m}(z_j)$.  The sum over $n$ goes from $n=0$
to $n=N-2$ since orbitals with $n \geq N-1$ are projected to zero.
Thus, up to an overall normalization, there are $N-2$ variational
parameters.

It is also possible to study other pairing symmetries within our approach.
These would yield states at different values of the flux $N_\phi$.  
Here, we focus on negative $p$-wave pairing, which appears most
consistent with previous numerical data 
\cite{Morf98,HaldaneRezayi00}.

The variational character of the wavefunctions we study
(Eq.~\ref{eq:PairedCFState}) implies that we need to optimize over
the set of parameters $\vec g\equiv( g_0, g_1, g_2, \ldots,
g_{N-2})$ to obtain a good trial wavefunction. The definition of a
``good" wavefunction is somewhat arbitrary, and one may attempt to
optimize various measures of its accuracy, e.g.\ the energy of the
wavefunction, the overlap with the exact groundstate, or the error
in the pair correlation function compared to the exact groundstate.
The chosen measures of accuracy are evaluated by Monte-Carlo, and
the variational parameters are then optimized.

It is instructive to verify that the Moore-Read state can be
reproduced as a CF-BCS state (Eq.~\ref{eq:PairedCFState}).
Numerically we find that for a suitable set of variational
parameters, $\vec g$, we are able to achieve overlaps in excess of
$0.99$ with the Moore-Read state for systems with up to 20
electrons. While this may seem a rather complicated reformulation of
the Moore-Read state, we can now perturb the wavefunction with our
variational parameters.

\begin{figure}[ttbp]
  \begin{center}
    \includegraphics[width=0.95\columnwidth]{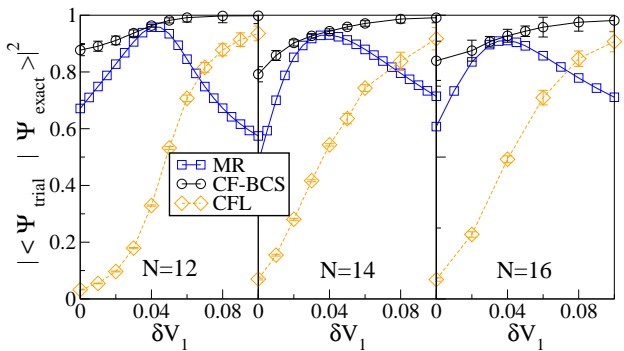}
  \end{center}
  \caption{  \label{fig:overlaps} (color online) Overlaps of trial
  states with exact groundstate as a function of the interaction
  parameter $\delta V_1$ for $N=12, 14$ and $16$ electrons.
  Here, $\delta V_1 = 0$ corresponds to the pure Coulomb interaction in
  the 1LL. The  optimized composite-fermionized BCS wavefunctions
  (Eq.~\ref{eq:PairedCFState}, black circles) have very high overlap except
  close to  $\delta V_1=0$, where the system
  is thought to be close to a phase transition. 
  The Moore-Read wavefunction (blue squares) is also
  good near $\delta V_1=0.04$, but falls off substantially at other
  values.  The composite fermion liquid wavefunctions (orange diamonds)
  are accurate at very high $\delta V_1$ only. Error bars indicate
  statistical errors where a Monte-Carlo algorithm was employed for 
  the evaluation of the overlaps. The high accuracy of the
  BCS wavefunctions over a broad range of interactions shows the
  large extent of the weak-pairing phase.
  }
\end{figure}

Fig.~\ref{fig:overlaps} shows overlaps between trial states and the
corresponding exact groundstates for different interactions
parametrized by the pseudopotential $\delta V_1$ (with $\delta V_1 =
0$ being the pure Coulomb interaction in the 1LL, and very large
$\delta V_1$ roughly corresponding to the interaction of the LLL).
We also show overlaps of the exact groundstate with the Moore-Read
state and the CF liquid (CFL) (here 
defined to be  Eq.~\ref{eq:PairedCFState} with the occupation
coefficients $g_n$ being unity below the chemical potential and zero
above). Results are shown for $N=12,14,16$. The dimensions of the
$L=0$ Hilbert space are respectively $d=52,291,2077$.  Although we
could in principle optimize over as many as $N-2$ variational
parameters, in practice we find very good wavefunctions using at
most the first $7$ parameters.  We remind the reader that the
variational parameters are included (like the $u$'s and $v$'s of BCS
theory) to optimize the shape of the pairing wavefunction.   Note
that the number of parameters used is far less than the dimension of
the $L=0$ Hilbert space, so the good agreement with exact
diagonalization is significant. Further we emphasize that the trial
states (Eq.~\ref{eq:PairedCFState}) have very high overlap with the
exact groundstate for a wide range of values of $\delta V_1$.

For a particular value of $\delta V_1 \approx 0.04$ the Moore-Read
state is also a very good trial state.   However even at this value
of $\delta V_1$, our trial states yield an improved representation
of the exact groundstate. For larger $\delta V_1$ the CF-BCS states
become even more accurate and can be continuously deformed into the
composite fermion liquid at large $\delta V_1$. As shown in
Fig.~\ref{fig:overlaps} when $\delta V_1$ gets close to zero (or
negative, not shown) the overlap drops. This behavior could be
expected considering prior work \cite{HaldaneRezayi00} showing a
nearby phase transition, as we discuss below.

\begin{figure}[ttbp]
  \begin{center}
    \includegraphics[width=0.95\columnwidth]{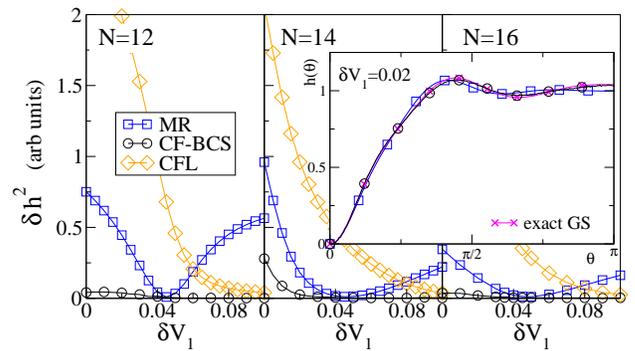}
  \end{center}
\caption{  \label{fig:paircorrelations} (color online) Squared error
$\delta h^2$
  in the pair correlation function of various trial wavefunctions
  compared with exact groundstates as a function of the interaction
  parameter $\delta V_1$ for $N=12, 14$ and $16$ electrons.   Symbols are as
  in Fig.~\ref{fig:overlaps}.  Again we find that composite-fermionized BCS wavefunctions
  are far more accurate than either Moore-Read or the CF liquid.
  Inset: Pair correlation functions $h(\theta)$ of the Moore-Read state and our trial wavefunction
  along with the exact groundstate at $N=14$ and $\delta V_1 = 0.02$.
  Our trial state is essentially indistinguishable from the exact
  groundstate, whereas the Moore-Read state is slightly different. }
\end{figure}

As mentioned above, overlaps are not the only possible measure of
the accuracy of a wavefunction.  In Fig.~\ref{fig:paircorrelations},
we show how the pair correlation functions $h(\theta)$ of different
trial states compare against those of the exact groundstate.  Here
$\theta$ is the angle between two particles on the sphere, and
we show the mean square error in the correlation function
$\delta h^2 = \int \mathrm{d}( \cos\!\theta) \, |h(\theta) -
h_\text{exact}(\theta)|^2$.   We note that if $\delta h^2=0$ for a
trial state, then it is identical to the exact groundstate (this
can be seen from the variational principle, noting that $h(\theta)$
fully determines the energy for 
a two-body interaction). Fig.~\ref{fig:paircorrelations} once more
illustrates that our trial wavefunctions are extremely accurate --
far more so than either the Moore-Read or CFL trial states. Again,
we find that near $\delta V_1=0$ our trial state fails to match the
exact pair correlation function to some extent.

In Fig.~\ref{fig:paircorrelations}, the trial wavefunctions have
been reoptimized with respect to  $\delta h^2$. The overlaps of
these new wavefunctions with the exact groundstate, would generally
be found to be slightly lower than those in Fig.~\ref{fig:overlaps},
but still remain very high (except in the vicinity of $\delta
V_1=0$).

Since, as mentioned above, we can continuously deform our
wavefunctions to have over 0.99 overlaps with the Moore-Read state,
we conclude that our CF-BCS wavefunctions are generally in the same
so-called weak-pairing phase as the Moore-Read state.  To further
emphasize this point, we note that a wavefunction in a weak-pairing
phase should have the property \cite{ReadGreen} that $g(\br) \sim
1/z$ at large distances $\br$.  While this is not obvious from the
form of Eq.~\ref{eq:PairedCFState} (particularly considering the
complexity of the projected wavefunctions $\tilde \phi$) we can
nonetheless establish it is true in several ways.  Firstly, we have
tried making the $1/z$ tail of the pair correlation function
explicit, writing $g(\br_i-\br_j)= a/(z_i-z_j) + f(\br_i-\br_j)$
before projection, decomposing only the function $f$ into orbitals
as in Eq.~\ref{eq:gFourier} and projecting these orbitals. We have
found that this procedure leads to equivalent results. Secondly, the
property $g(\br) \sim 1/z$ at large $\br$  implies that the $\bk\to
0$ orbitals are occupied with probability approaching
unity \cite{ReadGreen} (which would not be true of a strong pairing
phase).   It is easy for us to establish numerically that the lowest
orbitals ($n=0$) are indeed fully occupied unity by testing that 
increasing the value of the variational parameter $g_0$ does 
not change the wavefunction.

It is also worth checking that the exact groundstate is indeed
adiabatically connected to the Moore-Read state.  To this end, we
analyze the evolution of the energy gap for a family of Hamiltonians
that interpolate between the three-body contact interactions
$\mathcal{V}_3$, which yield the Moore-Read state as its exact
groundstate, and a two-body interaction Hamiltonian
$\mathcal{H}'_C$ corresponding to $\delta V_1=0.04$. In particular,
for any of the interactions $\mathcal{H}(x)=x \mathcal{V}^3 + (1-x)
\mathcal{H}'_C$, we find no indication that in the thermodynamic
limit the energy gap closes (data not shown). We conclude that the
exact groundstate of $\mathcal{H}'_C$ is adiabatically connected to
the Moore-Read state, confirming that the exact groundstate of
$\delta V_1 = 0.04$ is in the weak pairing phase \cite{ReadGreen}.

The main result of this work is the construction of a family of
accurate wavefunctions in the same topological phase as the
Moore-Read wavefunction.   This can be thought of as the 
composite fermionization of a weakly paired BCS wavefunction.   We
find that over a broad range of interactions these wavefunctions are
very accurate -- far more so than the Moore-Read wavefunction
itself, which should be thought of only as an example of a
wavefunction in a broad phase of matter. Indeed, the Moore-Read
state may be approximated extremely precisely by the form we
propose, and when doing so, the result does not particularly stand
out from other possible CF-BCS states. Although from a topological
standpoint, it is sufficient to identify the phase of matter,
from a practical standpoint, it is still 
valuable to have explicit forms of
wavefunctions \cite{Heinonen,JainKamilla97}, as this is important for
performing detailed calculations of excitation spectra and
other physical properties.  Although 
we have currently only analyzed the composite fermionization of
groundstate BCS wavefunctions, it is natural to consider a similar
procedure for excited states, which we will consider in future work.

Let us now discuss how our work reflects on and relates to
previously results. Prior work on the torus \cite{HaldaneRezayi00}
found a first order phase transition from a charge density wave
(CDW) state to a phase presumed to be the Moore-Read phase at
roughly the Coulomb point $\delta V_1=0$. This phase had the
required degeneracy of a weak pairing (Moore-Read) phase, but had
relatively low overlaps with the Moore-Read wavefunction itself.
Particle-hole symmetrizing the Moore-Read wavefunction increased the
overlap to 97\% for $N=10$ at one particular value of $\delta V_1$
but remained somewhat lower at other values.  We note that the
Moore-Read phase and its particle-hole conjugate are distinct phases
\cite{AntiPfaffian} and the effect of symmetrization is unclear and
remains a topic of current interest.   In our work on the sphere,
there is no possible mixing of states with their conjugates,
although we cannot determine whether a state or its conjugate would
occur in an experimental system.   On the sphere, it was previously
known \cite{Morf98} that the overlap of the exact groundstate with
the Moore-Read state has a peak at $\delta V_1 \approx 0.04$, and
also drops strongly near $\delta V_1=0$. However, on the sphere, it
was hard to distinguish the thermodynamic phase since there is no
groundstate degeneracy to use as a guide.  Our work, in contrast,
studies only trial wavefunctions in the Moore-Read phase.

In contrast to all prior work, our trial wavefunctions have high
overlaps over a very broad range of $\delta V_1$, confirming that
the weak pairing phase is robust to large changes in the
interaction. Our wavefunctions make a smooth transition between the
Moore-Read phase and the CF liquid at large $\delta V_1$.  It is
difficult to distinguish numerically if the groundstate of the LLL
still has some amount of pairing. To determine if at large $\delta
V_1$, the putative CFL still pairs (as previously suggested
\cite{HaldaneRezayi00}), a more careful study of the groundstate
for the LLL interactions would be required.  We note in passing that
the wide region of intermediate values of $\delta V_1$ (between
where the Moore-Read wavefunction is accurate and where the CFL
becomes accurate) which we describe extremely well with our
wavefunctions, could be hard to access with typical 2DEG samples but
could likely be realized using hole-doped samples \cite{Manfra} or
graphene \cite{Graphene}.

At interactions $\delta V_1 < 0.04$ our wavefunctions have
substantially better performance than the Moore-Read wavefunction.
However, near $\delta V_1 \approx 0$ our wavefunctions do not
perform as well as one might hope.  This is not surprising
considering that the groundstate of $\delta V_1 =0$ on the torus is
a CDW state \cite{HaldaneRezayi00}.  However, experiments, which see
a quantum Hall plateau, do not correspond to the pure Coulomb
interaction ($\delta V_1 = 0$) due to finite well width effects and
Landau-level mixing. It has also been noted \cite{HaldaneRezayi00}
that a more realistic interaction puts the physical system just
slightly on the quantum Hall side of the transition.  Indeed, it is
known experimentally \cite{StripeTiltField} that modifying the
electron interaction slightly by tilting the field pushes the system
from a quantum Hall state into a CDW state. Being that pure Coulomb
is thought to be on the other side of this phase transition, the
fact that our wavefunctions remain so good is perhaps surprising.
However, one might argue that since the CDW is frustrated by the
geometry of the sphere we can still match the groundstate
reasonably well with a sufficiently perturbed weak pairing
wavefunction, which remains adiabatically connected to the
Moore-Read state.

In Ref.~\cite{JainNoPfaffian} it was suggested that the gapped state
near the Coulomb point is best constructed within a composite
fermions basis without appeal to the Moore-Read wavefunction.  Our
wavefunction is indeed constructed in terms of composite fermions,
retains relatively high similarity with the exact groundstate, and
also remains adiabatically connected to Moore-Read. We believe this
result should put to rest concerns that the gapped phase near the
Coulomb point is not in the topological phase of the Moore-Read
state or its particle-hole conjugate \cite{AntiPfaffian}.

\acknowledgments

We are grateful to N.~Regnault for generously sharing data.
The authors acknowledge helpful discussions with E.~H.~Rezayi and
N.~R.~Cooper, and thank the Aspen Center for Physics for its
hospitality. G.M.~acknowledges support by EPSRC Grant
No.~GR/S61263/01.

\end{document}